# On the Use of Planetary Science Data for Studying Extrasolar Planets

*A science frontier white paper submitted to the Astronomy & Astrophysics 2020 Decadal Survey*

Thematic Area: Planetary Systems

Principal Author

Daniel J. Crichton
Jet Propulsion Laboratory, California Institute of Technology
Daniel.J.Crichton@jpl.caltech.edu
818-354-9155

Co-Authors:

J. Steve Hughes, Gael Roudier, Robert West, Jeffrey Jewell, Geoffrey Bryden, Mark Swain, T. Joseph W. Lazio (Jet Propulsion Laboratory, California Institute of Technology)

There is an opportunity to advance both solar system and extrasolar planetary studies that does not require the construction of new telescopes or new missions but better use and access to inter-disciplinary data sets. This approach leverages significant investment from NASA and international space agencies in exploring this solar system and using those discoveries as "ground truth" for the study of extrasolar planets. This white paper illustrates the potential, using phase curves and atmospheric modeling as specific examples. A key advance required to realize this potential is to enable seamless discovery and access within and between planetary science and astronomical data sets. Further, seamless data discovery and access also expands the availability of science, allowing researchers and students at a variety of institutions, equipped only with Internet access and a decent computer to conduct cutting-edge research.

## Scientific Context

Current and likely near-term frontiers in the characterization of extrasolar planets include topics such as understanding their demographics, determining their atmospheric compositions, assessing the role of clouds in their atmospheres, constraining their interior structures and properties, and investigating the interactions between their atmospheres or surfaces with the incident radiation and particle fluxes from their host stars. Moreover, these processes and properties may not be static over the life of a planet or its host star, but they may evolve with time, due to the evolution of the host star and potentially as a result of dynamic interactions between planets.

For the past several decades, there has been a comprehensive reconnaissance of our solar system. Ground- and space-based telescopes have monitored the appearance of solar system bodies over year- to decade-long time scales. Over the past five decades, NASA, the European Space Agency (ESA), and other international space agencies have invested billions of dollars in exploring the solar system, sending at least one spacecraft to every planet and to multiple dwarf planets and small bodies. Data returned from these spacecrafts include images, spectra, time series, field and wave measurements, and *in situ* samples.

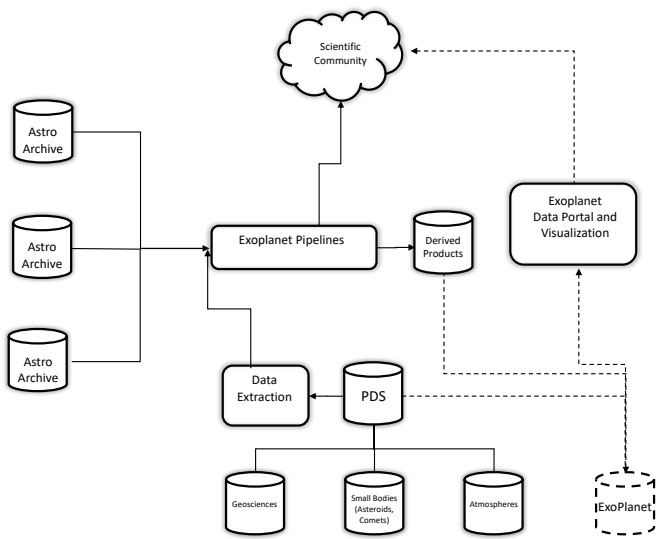

Figure 1. Vision for seamless discovery and data access between Astrophysical Archives and the Planetary Data System (PDS) enabling the use of solar system data as "ground truth" for input into extrasolar planetary data analysis.

The combination of these solar system and extrasolar planetary data sets already have enabled investigation of specific questions regarding the relation between the solar system and extrasolar planetary systems. Examples include the origin and delivery of water to the Earth and terrestrial planets by determining the deuterium/hydrogen (D/H) ratio of terrestrial planets, comets, and molecular clouds (e.g., van Dishoeck et al. 2014). Future questions that could be addressed by the combination of solar system and extrasolar planetary studies include

- **What is the range of interior structures that are obtained during planetary formation?** (E.g., combining interior structure knowledge of the Earth, Moon, Mars [InSight mission], Jupiter [Juno], and Saturn [*Cassini*] with constraints on extrasolar planetary bulk density and other measures of interior properties.)
- **What is and what determines the atmospheric composition of planetary atmospheres?** (E.g., combining data from the *Cassini* measurements of Saturn and the Juno measurements of Jupiter with transmission spectra of transiting extrasolar planets.)

Because of the history of how these various data have been gathered and the relatively recent discovery of extrasolar planets, the relevant data typically are stored in separate archives. Data for extrasolar planets are often contained in archives from ground-based telescopes funded by the NSF or in NASA's Astrophysics Archives whereas data from solar system planetary missions are stored in NASA's Planetary Data System (PDS). **Figure 1** presents a vision for seamless discovery and access between these various data archives that enables valuable "ground truth" from the solar system for the investigation of extrasolar planets and important context from extrasolar planetary systems about the planets of the solar system.

In the next two sections, we illustrate potential cross-disciplinary opportunities that make use of data between Astronomical Archives and the PDS. Because of space limitations, we illustrate only two cases, but other potential opportunities abound.

**Illustrative Opportunity: Planetary Phase Curves and Direct Imaging**

The evolution of a planet in time, its current atmospheric temperature, and its detectability in a direct imaging observation are determined, in part, by the ratio between the incident and reflected stellar flux. This ratio depends, in part, on the phase angle (star-planet-Earth angle). Conversely, when attempting to plan a mission or model the performance of a telescope for direct imaging, knowledge of the phase curve or the reflectivity as a function of the phase angle, as a function of wavelength, is crucial to predicting telescope or instrument performance accurately.

Ground-based studies suffer from a well-known limitation in determining the phase curve, particularly for the giant planets. Because the Earth's orbit is interior to all of the giant planets, the observed phase angles never exceed about 12°, and can be substantially less in the case of the ice giants. In order to use solar system planets as analogs for extrasolar planets, the full range of phase angles should be known (**Figure 2**).

NASA's missions to the outer solar system have provided phase curve measurements for Jupiter and Saturn, and at least limited measurements for Uranus and Neptune. These missions include the Pioneer 10 and 11, Voyager 1 and 2, *Galileo*, *Cassini*, and the current Juno mission; the number and diversity of measurements will likely increase over the next decade due to the Lucy mission (planned launch 2021 October), the Europa Clipper mission, ESA's JUpiter ICy Moon Exploration (JUICE) as well as potential future missions such as an Ice Giant Mission to either Uranus or Neptune or both. However, the data from which phase curves can be constructed have been acquired by Planetary Science missions and are typically stored in NASA's Planetary Data System (PDS), rather than in NASA's Astrophysics Archives.

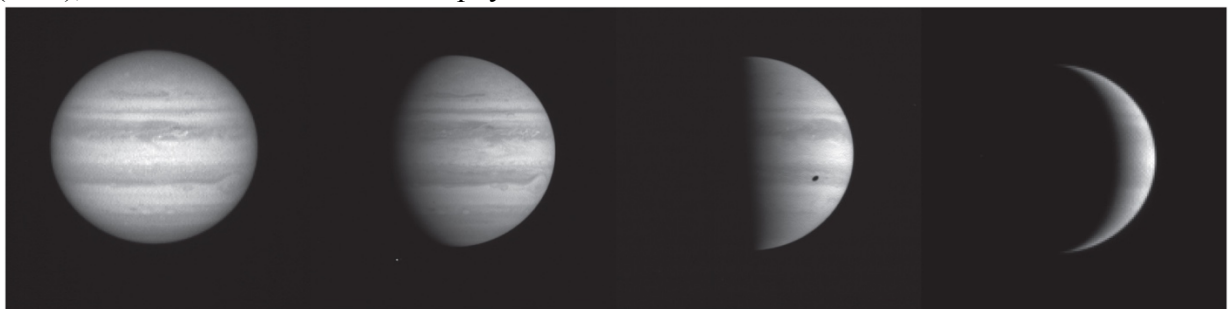

Figure 2. Future extrasolar planet direct imaging observations will view extrasolar planets at a variety of phase angles. Both telescope planning and the interpretation of future extrasolar planet images will benefit from access to images of solar system planets acquired at a variety of phase angles by Planetary Science missions. Shown here are images of Jupiter at different phase angles acquired by the *Cassini*/Imaging Science Subsystem. Increased inter-operability between NASA's Planetary Data System (PDS) and Astrophysics Archives would facilitate both telescope design planning for extrasolar planet observations and subsequent data interpretation. (From Mayorga et al. 2016)

**Illustrative Opportunity: Reflected Light and Extrasolar Planetary Phase Curves**

Phase curve analyses—unresolved stellar and planetary spectroscopic observations over a full orbital revolution—probe the content, structures, and dynamics of extrasolar planetary atmospheres, particularly in combination with transit spectroscopy. A state-of-the-art example for hot Jupiters is the WASP-43b phase curve analyses of Stevenson et al. (2017). The next generation of datasets (*JWST*, ARIEL) combined with the community's increasing focus on colder targets such as super Earths and warm Neptunes will push the ongoing theoretical effort into more detailed and complex modeling and unveil a regime of exoplanetary atmospheric physics that has been poorly explored, due to the current observational bias towards high temperature atmospheres of gas giants. In the near-IR, the relative contribution (dominance) of the thermal emission from the planet compared to the stellar light bouncing through the extrasolar planet's atmosphere will change, requiring the modeling to consider gas absorbers in combination with scattering and reflecting atmospheric components such as aerosols and clouds.

Jupiter represents an extreme case of this radiative transfer regime in the UV and NIR. As a cold gas giant, both scattered and reflected light contribute to the Jovian signal, in comparison to the current set of extrasolar planets for which atmospheric modeling has been developed. In order to validate a phase curve modeling tool for extrasolar planets, Zhang et al. (2013) stress-tested a new forward model and Bayesian retrieval facility against *Cassini*/ISS images, which have much higher signal-to-noise ratios relative to extrasolar planetary data. Zhang et al. (2013), and subsequent efforts, modeled the Jovian atmosphere by building on the state-of-art atmospheric parametrization and augmenting it with a separate and distinctive tropospheric haze layer and clouds as opposed to a phenomenological parametrization of the bundled {cloud + tropospheric haze + gas content} at the bottom of the Jovian atmosphere.

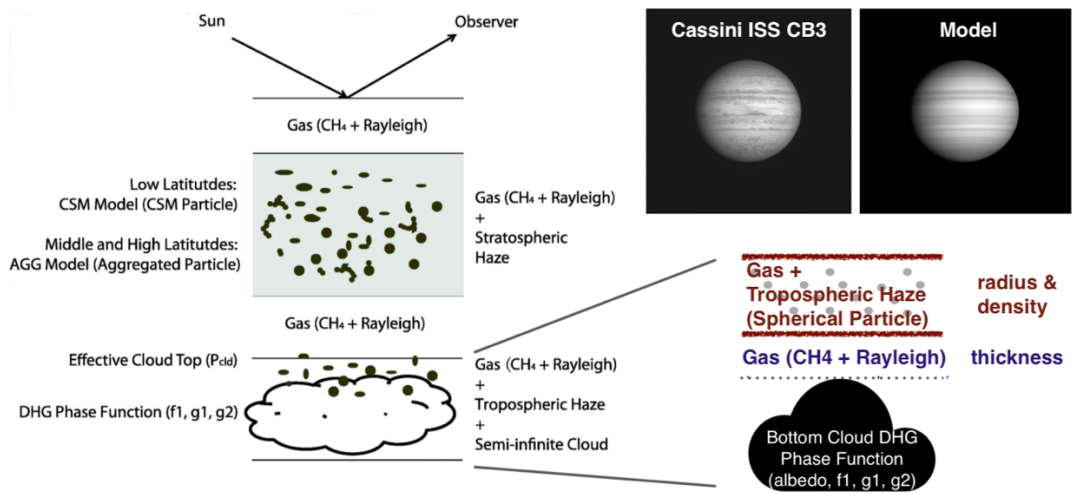

Figure 3: (*left*) Illustration of the state-of-the-art model used to interpret a *Cassini*/ISS dataset for Jupiter. (From Zhang et al. 2013.) (r*ight*) Augmentation to the previous model at the bottom layer of the atmosphere. Additional parameters are tropospheric haze radius, density, sub-tropospheric gas layer thickness, and cloud parameters for each latitude band as well as the cloud albedo for each considered wavelength.

**Figure 3** illustrates the parametrization of the Jovian atmosphere, and a sample comparison between this new modeling approach and *Cassini*/ISS data at low phase angle, for which the Jovian disc is almost fully probed. While the complexity of the Jovian parametrization will transpose only partially to extrasolar planets, modeling aerosols and clouds accurately will be decisive in interpreting phase curves from *JWST* and ARIEL/CASE. For each phase angle or extrasolar planet orbital phase, a forward modeled atmosphere will be contracted into a single data point, averaging over latitudinal variations.

**Key Advance: Integrated Astronomical-Planetary Sciences Data Access and Analytics**

As has been illustrated above, advances could be enabled not only by new missions or telescopes, but by seamless discovery and access to relevant data in both Astrophysics and Planetary Science Archives. Over the last decade, there has been tremendous growth of data across all disciplines providing an opportunity for driving new analysis, particularly by bringing data together across disciplines, missions, telescopes, and instruments. Advances in data-driven techniques offer opportunities for significantly leveraging the data and computational capabilities by shifting from human-intensive analysis approaches to automated data mining techniques (e.g., machine learning) to helping to automate extraction and key features in data.

Significant advances have been made to enable well-curated data archives, in which the data are annotated and structured as well as services are provided to search and extract data. While many archives have focused traditionally on preservation and distribution of data, there is an increasing opportunity to construct analytical data pipelines that can take archive data and process them through computational pipelines and workflows that can be used to transform and combine data. This is also an opportunity for integrating the data and the computation, which can be used to extract and classify features that may be useful for additional subsequent analysis.

To date, much of the focus of this work has been on the data within a single archive, or data within the archives of a single discipline (e.g., Earth Sciences). Indeed, within some disciplines, such capabilities are becoming routine. Over the next decade, there is the opportunity to begin to integrate distinct disciplines such as Astronomy & Astrophysics with Planetary Sciences, particularly in the area of extrasolar planets (**Figure 1**).

Not only could there be integrated data access, there is the opportunity to enable computation to enable different types of analytics. In addition to increasing automation for data analysis and data understanding, new approaches for human-machine interfaces could become integrated with analytical capabilities to enable *interactive analytics*. This will allow for more data exploitation, with computation and on-demand processing as part of the process of interrogating data through visual tools in order to better scale data analysis for massive data sets. **Figure 1** shows the opportunity for bringing together a set of data services to support exoplanet research across astronomy and planetary sciences by constructing a virtual data environment to access and analyze data across these disciplines.

There are a variety of ways in which extrasolar planets could be used to bridge the data archives of Astronomy & Astrophysics and Planetary Sciences. One approach would be for Astronomy & Astrophysics archives to incorporate, for relevant data, the PDS4 Information Model, which the PDS developed in order to enable interoperability and provide a foundation for interdisciplinary science across the various planetary science disciplines at an international scale. In general, information models allow humans and computers to "communicate" about digital content and to interoperate across science disciplines through common terms and definitions. In addition, an information model is leveraged as a set of information requirements that help drive software and services development to both steward and discover the data. In the specific case of Planetary

Science, the PDS4 Information Model is the community's controlled vocabulary and associated semantics to capture, manage, discover, and use the data.

The PDS4 Information Model enables interdisciplinary science through extensions into new and related disciplines, such as Astronomy & Astrophysics. The PDS4 Information Model's object-oriented modeling paradigm and the multi-level governance scheme results in a modeling hierarchy with one common, several discipline-specific, and an expanding set of local or project sub-models. Extensions leverage common terms and definitions while allowing new specificity as required.

The PDS4 Information Model can be extended to support provenance information, the historical record of the inputs, entities, systems, and processes used to produce a data product. The "data processing event" pattern, an extension of the W3C PROV Ontology Recommendation (PROV-O) is being added to the information model and extended to provide a simple template that relates the fundamental aspects of a science data processing event such as its timing, inputs, outputs and causative agents. In addition to providing a critical component of a trusted data archive, the provenance extension will also be used to promote reproducibility, one component of the precision of a measurement.

We close with three other notes on these opportunities.

1. We have illustrated particular science uses cases, focusing on extrasolar planets. Other potential opportunities abound, such as identifying asteroids and Kuiper Belt objects (of interest to the PDS Small Bodies Node) within astronomical surveys.

2. We have described how the PDS4 Information Model could be used as the basis for a bridge between the two disciplines. Other potential information models could also be adopted or used.

3. Our focus in this white paper has been on the scientific context and potential advances over the next decade, if data about the planets in our own solar system can be easily discovered and accessed. A complementary aspect of data discovery and access is that it can broaden participation. Accessing a data archive requires little more than a computer, an Internet connection, and some amount of data storage. These resources are well within the capabilities of a broad range of institutions and even individuals. A consequence of enabling data discovery and access is that doing so permits researchers and students at a variety of institutions to participate in planetary and extrasolar planetary discovery and studies.